\documentclass[letterpaper,12pt,notoc]{JHEP3}
\def\pd{\partial}
\def\Slash{{\!\!\!\!/}}

\usepackage{graphicx}
\usepackage{amsmath}
\usepackage{amssymb}

\preprint{ \hbox{}\hfill arXiv:1012.4953}

\title{Two dimensional RG flows and Yang-Mills instantons}

\author{Edi Gava$^a$, Parinya Karndumri$^{a,\, b}$ and K. S. Narain$^c$\\
$^a$INFN, Sezione di Trieste, Italy\\
$^b$International School for Advanced Studies (SISSA), via Bonomea
265, 34136 Trieste, Italy \\
$^c$The Abdus Salam International Centre for Theoretical Physics,
Strada Costiera 11, 34100 Trieste, Italy \\
E-mail: \email{gava$@$ictp.it}, \email{karndumr$@$sissa.it},
\email{narain$@$ictp.it}}

\abstract{We study  RG flow solutions in (1,0) six dimensional
supergravity coupled to an anti-symmetric tensor and  Yang-Mills
multiplets corresponding to a semisimple group $G$. We turn on $G$
instanton gauge fields, with instanton number $N$, in the
conformally flat part of the 6D metric. The solution interpolates
between two (4,0) supersymmetric $AdS_3\times S^3$ backgrounds with
two different values of $AdS_3$ and $S^3$ radii and
 describes an RG flow in the dual 2D  SCFT. For the single
instanton case and $G=SU(2)$, there exist a consistent reduction
ansatz to three dimensions, and the solution in this case can be
interpreted as an uplifted 3D solution.  Correspondingly, we present
the solution in the framework of $N=4$ $(SU(2)\ltimes
\mathbf{R}^3)^2$ three dimensional gauged supergravity. The flows
studied here are of  v.e.v. type, driven by a vacuum expectation
value of a (not exactly) marginal operator of dimension two in the
UV. We give an interpretation of the supergravity solution in terms
of the D1/D5 system in type I string theory on K3, whose effective
field theory is expected to flow to a (4,0) SCFT in the infrared.}

\begin{document}
%%%%%%%%%%%%%%%%%%%%%%%%%%%%%%%%%%%%%%%%%%%%%%%%%%%%%%%%%%%%%%%%%%%%%%%
\section{Introduction}
Renormalization Group flows have been a subject of much study soon
after the proposal of the AdS/CFT correspondence \cite{maldacena}.
For supersymmetric flows, solutions interpolating between two AdS
critical points of the dual supergravity theories can be obtained by
looking at BPS equations. The flows describe typically deformations
of the CFT at the UV fixed point by relevant operators or, less
typically, by vacuum expectation values of relevant operators and,
like in the present case, by vevs of marginal operators. The theory
is then driven to another fixed point in the IR. These flow
solutions have been originally studied in the context of five
dimensional or three dimensional gauged supergravities to describe
RG flows in the dual four dimensional or two dimensional SCFT's.
Examples of these can be found in \cite{fgpw, an, gir} for the 4D
case and  in \cite{bs,gkn, AP} for the 2D case. In some cases, these
solutions can be uplifted to higher dimensions by using consistent
truncation formulae, and this will also be the case for a class of
solutions of the present paper.
\\
\indent Here  we study some RG flow solutions in the context of
AdS$_3$/CFT$_2$. Our starting point is (1,0) six dimensional
supergravity coupled to a tensor multiplet and to $SU(2)$ Yang-Mills
multiplets. In an earlier paper \cite{KK}, we have shown that the
$SU(2)$ reduction of this theory gives rise to $N=4$ three
dimensional gauged supergravity with scalar manifold
$SO(4,4)/SO(4)\times SO(4)$. We first discuss a 6D flow solution
which, in fact, is the lift to 6D of an RG flow of the 3D $N=4$
gauged supergravity and preserves half of the supersymmetries. This
solution involves an $SU(2)$ instanton on ${\bf R}\times S^3$, ${\bf
R}$  being the radial coordinate, with topological charge equal to
1, which in the 3D setting is seen as a scalar's background. The
instanton interpolates between the $|0\rangle$ Yang-Mills vacuum
with winding number 0 in the IR and $|1\rangle$ vacuum with winding
number 1 in the UV. We then move to study solutions involving
multi-instanton gauge fields of an arbitrary semisimple gauge group
$G$. In this case, the solution is genuinely six dimensional in the
sense that it cannot be obtained as an uplifted solution of a three
dimensional theory, roughly, because it involves higher modes on
$S^3$. The instanton interpolates between $|N\rangle$ vacuum in the
UV and $|0\rangle$ in the IR. The solution has been studied long ago
in \cite{duff, Pope_instanton_moduli} but in different contexts. In
this paper, we will look at it from another point of view by
regarding it as an RG flow solution interpolating between the UV and
IR CFT's corresponding to two AdS$_3$ limits. The central charge at
the two fixed points of course respects the c-theorem and admits an
interpretation in terms of the dynamics of the $D1/D5$ dual system
giving rise to a $(4,0)$ SCFT in the decoupling limit\cite{D1D5I,
Douglas, witten40}. What emerges from our solutions is that the flow
from UV to IR is essentially a manifestation of the Coulomb branch,
where $N$ D5-branes, if the instanton number is $N$, decouple from
the D1D5 system as we move towards the asymptotic infrared region.
Therefore in the IR, the central charge, which is linear in the
number of D5 branes, decreases accordingly. We also discuss the
contribution of the gauge Chern-Simons terms to the central charge
of $SU(2)$ left moving R-symmetry current algebra. In doing this, we
will give a different derivation of the result obtained in
\cite{Brown Henneaux}: our derivation involves the computation of
chiral correlators of the stress energy tensors or of the currents,
respectively, in the spirit of \cite{Witten}. As a byproduct, we
will obtain a holographic derivation of Virasoro's and current
algebra Ward identities.
\\ \indent The paper is organized as follows. We start by finding an RG flow solution in
the (1,0) six dimensional supergravity coupled to an anti-symmetric
tensor and $SU(2)$ Yang-Mills multiplets in section \ref{6Dflow}. As
we mentioned, there exist a consistent reduction to three
dimensional gauged supergravity for the $SU(2)$ gauge fields
describing a single instanton. In appendix \ref{3Dflow}, we also
give the same solution in the framework of $N=4$ three dimensional
gauged supergravity. In section \ref{instanton}, we generalize the
solution found in section \ref{6Dflow} to the case in which the
gauge fields of an arbitrary semisimple gauge group describe a
configuration of $N$ instantons. We also give an interpretation of
the solution in term of D1/D5 brane system in type I string theory.
We then discuss the central charge of the dual CFT. As mentioned
above, we rederive Virasoro and current algebra central charges, and
give some details of the derivation in appendix \ref{derivation}. We
end this paper by making some conclusions and comments in section
\ref{conclusion}.
%%%%%%%%%%%%%%%%%%%%%%%%%%%%%%%%%%%%%%%%%%%%%%%%%%%%%%%%%%%%%%%%%%%%%%%%%%%%%%%%%%%%%%%%%%%%%%%%%%%%%%%%%%%%%%%%%%%%%
\section{An RG flow solution from six dimensional supergravity on $SU(2)$ group manifold}\label{6Dflow}
In this section, we study an RG flow solution of the (1,0) six
dimensional supergravity coupled to an anti-symmetric tensor and
$SU(2)$ Yang-Mills multiplets. This is a special case, where
$G=SU(2)$, of the theory studied in \cite{KK} in the context of
$SU(2)$ reduction. This theory is an ungauged version of the (1,0)
six dimensional gauged supergravity constructed in \cite{nishino1,
nishino}. Here, we will find a flow solution in the six dimensional
framework. At the fixed points, the solution can be interpreted as
supersymmetric $AdS_3\times S^3$ backgrounds.\\ \indent We start by
giving some necessary formulae we will use in this and the following
sections. Six dimensional supergravity coupled to an antisymmetric
tensor multiplet admits a Lagrangian formulation. According to the
formulation in \cite{nishino}, the Bianchi identity for the three
form field $\hat{G}_3$ is modified in the presence of gauge fields
$\hat{A}$ by
\begin{equation}
\hat{d}\hat{G}_3=v^z \textrm{Tr}(\hat{F}\wedge \hat{F}).
\end{equation}
The equation of motion for $\hat{G}_3$ is also modified to
\begin{equation}
\hat{d}(e^{2\hat{\theta}}\hat{*}\hat{G}_3)=-\tilde{v}^z\textrm{Tr}(\hat{F}\wedge
\hat{F}).
\end{equation}
where $\hat{\theta}$ is the scalar field in the tensor multiplet. It
has been shown in \cite{nishino} that when one of the parameters
$v^z$ and $\tilde{v}^z$ vanishes, an invariant Lagrangian can be
written down. In \cite{KK}, we have chosen $v^z=1$ and
$\tilde{v}^z=0$ and shown that there is a consistent $SU(2)$
reduction to $N=4$ three dimensional gauged supergravity. Throughout
this section, we will work with this choice so that the solution in
this section can be considered as a flow solution of the
$SU(2)\times SU(2)\sim SO(3)\times SO(3)$ three dimensional gauged
supergravity. Furthermore, we will use $SU(2)$ and $SO(3)$
interchangeably in this paper.
\\ \indent In order to find supersymmetric solutions, we
need supersymmetry transformations for fermionic fields. We only
give here the expressions and refer the readers to \cite{nishino1,
nishino} for more details. We distinguish six and three dimensional
fields by putting a hat on all six dimensional fields. The
supersymmetry transformations of the gravitino $\psi_\mu$, gauginos
$\lambda^I$ and the fermion in the tensor multiplet $\chi$ are given
by \cite{nishino}
\begin{eqnarray}
\delta \psi_M &=& \hat{D}_M \epsilon +\frac{1}{24}e^{\hat{\theta}}
\Gamma^{NPQ}\Gamma_M \hat{G}_{3NPQ}\epsilon, \label{deltapsi}\\
\delta \lambda^I&=&\frac{1}{4}\Gamma^{MN}\hat{F}^I_{MN}\epsilon,
\label{deltalambda}
\\
\delta \chi&=&\frac{1}{2}\Gamma^M\partial_M\hat{\theta}\epsilon
-\frac{1}{12}e^{\hat{\theta}}\Gamma^{MNP}\hat{G}_{3MNP}\epsilon
\label{deltachi}
\end{eqnarray}
where we have given only the bosonic contributions. Note that we
have not put a hat on fermions because we will not consider three
dimensional fermions here. This is all we need for finding BPS
solutions.
\\ \indent We now review the reduction ansatz used in \cite{KK}
\begin{eqnarray}
d\hat{s}^2&=&e^{2f}ds^2+e^{2g}h_{\alpha \beta}\nu^\alpha \nu^\beta
\, , \nonumber \\ \hat{A}^I&=&A^I+A^I_\alpha \nu^\alpha, \qquad
\nu^\alpha=\sigma^\alpha-g_1A^\alpha \, ,\nonumber \\
\hat{F}^I&=&d\hat{A}^I+\frac{1}{2}g_2f_{IJK}\hat{A}^J\wedge\hat{A}^K\nonumber
\\ &=& F^I-g_1A^I_\alpha F^\alpha +\mathcal{D}A^I_\alpha\wedge
\nu^\alpha + \frac{1}{2}(g_2A^J_{\alpha} A^K_{\beta}
f_{IJK}-\epsilon_{\alpha \beta \gamma}A^I_\gamma)\nu^\alpha \wedge
\nu^\beta, \nonumber \\
\hat{G}_3&=&\tilde{h}\varepsilon_3+\bar{F}^\alpha\wedge \nu^\alpha
+\frac{1}{2}K_{\alpha \beta}\wedge\nu^\alpha \wedge\nu^\beta
+\frac{1}{6}\tilde{a}\epsilon_{\alpha\beta\gamma}\nu^\alpha\wedge\nu^\beta\wedge\nu^\gamma\label{ansatz0}
\end{eqnarray}
where
\begin{eqnarray}
\tilde{h}&=&h\varepsilon_3+\tilde{F}^I\wedge
A^I-\frac{1}{6}g_2A^I\wedge A^J\wedge A^Kf_{IJK},\nonumber \\
\bar{F}^\alpha &=&A^I_\alpha (2F^I-g_1A^I_\alpha F^\alpha)-6ag_1
F^\alpha, \nonumber \\
\tilde{a}&=&6a-A^I_\alpha A^I_\alpha+\frac{1}{3}g_2A^I_\alpha
A^J_\beta A^K_\gamma f_{IJK}\epsilon_{\alpha\beta\gamma}, \nonumber
\\
\textrm{and} \qquad K_{\alpha
\beta}&=&A^I_\beta\mathcal{D}A^I_\alpha-A^I_\alpha\mathcal{D}A^I_\beta\,
.
\end{eqnarray}
In the present case, indices $I,J,K=1,2,3$ are $SU(2)$ adjoint
indices. The structure constant $f_{IJK}$ will be replaced by
$SU(2)$ structure constant $\epsilon_{IJK}$. $\mathcal{D}$ is the
$SU(2)\times SU(2)$ covariant derivative with coupling constants
$g_1$ and $g_2$, respectively,
\begin{equation}
\mathcal{D}A^I_\alpha=dA^I_\alpha +g_1\epsilon_{\alpha
\beta\gamma}A^\beta A^I_\gamma+g_2\epsilon_{IJK}A^JA^K_\alpha\, .
\end{equation}
\indent In order to obtain the vielbein on the $S^3$ part, we
introduce the scalar matrix $L^i_\alpha$ defined by
\begin{equation}
h_{\alpha\beta}=L^i_\alpha L^i_\beta\, .
\end{equation}
The left invariant $SU(2)$ one forms satisfy
\begin{equation}
d\sigma^\alpha =-\frac{1}{2}\epsilon_{\alpha\beta\gamma}\sigma^\beta
\wedge\sigma^\gamma\, .
\end{equation}
Finally, there is a truncation in both bosonic and fermionic fields
in the reduction of \cite{KK}. The relevant relation for our
discussion is the bosonic one given by
\begin{equation}
h_{\alpha\beta}=e^{\theta-2g}(12a\delta_{\alpha \beta}-2A^I_\alpha
A^I_\beta)\label{truncation_relation}
\end{equation}
where $a$ is a constant. This relation is a consequence of
truncating out massive vector fields in three dimensions.\\ \indent
In the ansatz for the flow solution, we set
$A^I_\mu=A^\alpha_\mu=0$, $A^I_\alpha=\delta^I_\alpha A$ and
$L^i_\alpha=\delta^i_\alpha$. After various scalings discussed in
\cite{KK} together with the standard domain wall ansatz in three
dimensions, we find
\begin{eqnarray}
ds^2&=&e^{2w(r)}(-dx^2_0+dx_1^2)+dr^2+\frac{e^{2q(r)}}{4g_1^2}\delta_{\alpha\beta}\sigma^\alpha\sigma^\beta,
\nonumber \\
\hat{A}^I&=&A^I_\alpha \sigma^\alpha =A\delta^I_\alpha\sigma^\alpha=A\sigma^I, \nonumber \\
\hat{F}^I&=&\frac{1}{g_1}dA\wedge
\sigma^I+\frac{1}{2g_1^2}(g_2A^2-g_1A)\epsilon_{IJK}\sigma^J\wedge\sigma^K, \nonumber \\
\hat{G}_3&=&h\varepsilon_3+\frac{1}{6g_1^3}\bigg(g_1+2g_2A^3-3g_1A^2\bigg)\epsilon_{IJK}\sigma^I\wedge\sigma^J\wedge\sigma^K\,
. \label{ansatz}
\end{eqnarray}
The scalings have been performed to restore the $g_1$ and $g_2$ in
the appropriate positions in the solution. This makes the comparison
with the three dimensional solution given in appendix \ref{3Dflow}
clearer. Notice the particular ansatz for $A^I_\alpha$ which gives
$K_{\alpha\beta}=0$. This is the reason for the consistency of the
truncation of the three dimensional gauge fields, $A^I_\mu=0$ and
$A^\alpha_\mu=0$. It can be easily checked that all the three
dimensional field equations given in \cite{KK} are satisfied by our
ansatz. The $S^3$ part of the metric and that in \eqref{ansatz0} are
related by
\begin{equation}
\frac{e^{2q}}{4g_1^2}\delta_{\alpha\beta}=e^{2g}h_{\alpha\beta}=2e^\theta
(1-A^2)\delta_{\alpha\beta}\label{metric mathing}
\end{equation}
where we have used \eqref{truncation_relation} after scalings. We
will see later that our solution satisfies this relation and is
indeed a solution of the theory obtained in \cite{KK}. The
supersymmetric flow solution can be found by considering the Killing
spinor equations coming from the supersymmetry transformation of
fermions. From the metric, we can read off the vielbeins
\begin{equation}
\hat{e}^a=e^wdx^\mu,\,\,\, \hat{e}^{\hat{r}}=dr,\,\,\, \textrm{and}
\,\,\, \hat{e}^i=\frac{e^q}{2g_1}\sigma^i
\end{equation}
We can compute the following spin connections
\begin{eqnarray}
\hat{\omega}^{a\hat{r}}&=&w'\hat{e}^a,\nonumber\\
\hat{\omega}^{\hat{r}i}&=&-q'\hat{e}^i,\nonumber\\
\hat{\omega}^{ij}&=&-g_1e^{-q}\epsilon_{ijk}\hat{e}^k\,
.\label{spinconnection}
\end{eqnarray}
The index $a$ is the tangent space index for $\mu=0,1$, and $'$
means $\frac{d}{dr}$. For conveniences, we also repeat here the
decompositions of the six dimensional gamma matrices from
\cite{PopeSU2}
\begin{eqnarray}
\Gamma^{\hat{A}}&=&(\Gamma^a,\Gamma^i), \qquad \Gamma^a=\gamma^a
\otimes
\mathbb{I}_2 \otimes \sigma_1, \nonumber \\
\Gamma^i&=& \mathbb{I}_2 \otimes \gamma^i\otimes \sigma_2, \qquad
\Gamma_7=\mathbb{I}_2 \otimes \mathbb{I}_2 \otimes \sigma_3,
\nonumber \\
\gamma^{abc}&=&\epsilon^{abc},\qquad \gamma^{ijk}=i\epsilon^{ijk},
\qquad \{\gamma_a,\gamma_b\}=2\eta_{ab},\, \,\,
\{\gamma_i,\gamma_j\}=2\delta_{ij}
\end{eqnarray}
with the same conventions given in \cite{PopeSU2}. We further
specify the three dimensional gamma matrices by the following choice
\begin{equation}
\gamma^0=i\tilde{\sigma}_2,\qquad \gamma_1=\tilde{\sigma}_1, \qquad
\gamma^2=\tilde{\sigma}_3\, .
\end{equation}
\indent Using \eqref{deltapsi}, \eqref{deltalambda} and
\eqref{deltachi}, we find
\begin{eqnarray}
\delta\lambda^I=0&:&\, A'=-2(g_2A^2-g_1A)e^{-q},\nonumber \\
\delta\chi =0&:& \,
\theta'=e^\theta\big[h-8e^{-3q}\big(g_1+2g_2A^3-3g_1A^2\big)\big],\nonumber \\
\delta\psi_i =0&:& \,
q'=-g_1e^{-q}+\frac{1}{2}\big[e^\theta\big[h+8e^{-3q}\big(g_1+2g_2A^3-3g_1A^2\big)\big]\big],\nonumber
\\
\delta \psi_a=0&:&\,
w'=-\frac{1}{2}e^\theta[h+8e^{-3q}(g_1+2g_2A^3-3g_1A^2)]
\end{eqnarray}
where we have used
$\tilde{\sigma}_3\otimes\mathbb{I}_2\otimes\mathbb{I}_2\epsilon=\epsilon$.
So, the solution preserves half of the (1,0) supersymmetry in six
dimensions. As in \cite{KK}, we fix $h$ by using the equation of
motion for $\hat{G}_3$
\begin{equation}
\hat{D}(e^{2\hat{\theta}}\hat{*}\hat{G}_3)=0\, .
\end{equation}
This gives $he^{3q+2\theta}=c_1$ with a constant $c_1$. Using this
result and changing the coordinate $r$ to $\tilde{r}$ given by
$\frac{d\tilde{r}}{dr}=e^{-q}$, we find that the above equations can
be rewritten as
\begin{eqnarray}
\theta'&=&e^{\theta-2q}(c_1e^{-2\theta}-8\tilde{a})\, ,\label{theta}\\
q'&=&-g_1+\frac{1}{2}e^{\theta-2q}(e^{-2\theta}c_1+8\tilde{a})\, ,\label{g}\\
w'&=&-\frac{1}{2}e^{\theta-2q}(c_1e^{-2\theta}+8\tilde{a})\, ,\label{f}\\
A'&=&-2(g_2A^2-g_1A) , \label{A}
\end{eqnarray}
where $\tilde{a}=g_1+2g_2A^3-3g_1A^2$. The $'$ is now
$\frac{d}{d\tilde{r}}$. Before solving these equations, let us look
at the fixed points given by the conditions $\theta'=q'=A'=0$. There
are two fixed points:
\begin{itemize}
  \item I:
  \begin{eqnarray}
  A&=&0, \qquad \theta =\frac{1}{2}\ln{\frac{c_1}{8g_1}},\nonumber \\
  q&=&\frac{1}{4}\ln{\frac{8c_1}{g_1}},
  \end{eqnarray}
  \item II:
  \begin{eqnarray}
  A&=&\frac{g_1}{g_2}, \qquad \theta =\frac{1}{2}\ln{\frac{c_1g_2^2}{8g_1(g_2^2-g_1^2)}},\nonumber \\
  q&=&\frac{1}{4}\ln{\frac{8c_1(g_2^2-g_1^2)}{g_1g_2^2}}.
  \end{eqnarray}
\end{itemize}
Equation \eqref{A} can be solved and gives
\begin{equation}
A=\frac{g_1}{g_2-e^{g_1C_2-2g_1\tilde{r}}}.\label{Asol}
\end{equation}
Taking the combination \eqref{theta}+ 2 \eqref{g}, we find
\begin{equation}
z'=2e^{-z}c_1-2g_1\label{zeq}\\
\end{equation}
where $z=\theta+2q$. From \eqref{zeq}, we find the solution for $z$
is
\begin{equation}
z=\ln \frac{c_1-e^{-2g_1\tilde{r}+C_3}}{g_1}\label{zsol}.
\end{equation}
From \eqref{Asol}, we see that the fixed point I is at
$\tilde{r}\rightarrow -\infty$ while the II point is at
$\tilde{r}\rightarrow \infty$. Regularity of $A$ requires that
$-e^{g_1C_2}$ must have the same sign as $g_2$. For convenience, we
choose
\begin{displaymath}
C_2=\frac{1}{g_1}\ln (-g_2).
\end{displaymath}
From \eqref{zsol}, $z$ blows up as $\tilde{r}\rightarrow -\infty$,
so the solution breaks down at the I point. To overcome this
problem, we choose $z$ to be constant in such a way that \eqref{zeq}
is satisfied identically. This can be achieved by setting
\begin{equation}
z=\ln{\frac{c_1}{g_1}}.
\end{equation}
This means $\theta=\ln\frac{c_1}{g_1}-2q$. We can see that this
condition is satisfied at both fixed points, and equations \eqref{g}
and \eqref{theta} collapse to a single equation namely
\begin{equation}
q'=4e^{-4q}\frac{c_1}{g_1}\Big(g_1-\frac{e^{4g_1\tilde{r}}(3+e^{2g_1\tilde{r}})g_1^3}{g_2^2(1+e^{2g_1\tilde{r}})^3}\Big)-\frac{g_1}{2}.
\end{equation}
This equation can be solved, and we find
\begin{equation}
q=\frac{1}{4}\ln
\bigg[8e^{-2g_1\tilde{r}}\bigg(c_1e^{2g_1\tilde{r}}\bigg(\frac{1}{g_1}-\frac{g_1}{g_2^2}\bigg)-\frac{c_1g_1(2+3e^{2g_1\tilde{r}})}{g_2^2(1+
e^{2g_1\tilde{r}})^2} +\frac{54C_4}{g_2^2}\bigg)\bigg].
\end{equation}
In order to make the solution for $q$ interpolates between the two
values at both fixed points, we need to choose
\begin{equation}
C_4=\frac{c_1g_1}{27}.
\end{equation}
We finally find
\begin{eqnarray}
A&=&\frac{g_1}{g_2(1+e^{-2g_1\tilde{r}})}\label{Asolfinall} \\
q&=& \frac{1}{4}\ln
\frac{8c_1(g_2^2+2g_2^2e^{2g_1\tilde{r}}+(g_2^2-g_1^2)e^{4g_1\tilde{r}})}
{g_1g_2^2(1+e^{2g_1\tilde{r}})^2} \label{gsolfinal} \\
w&=&-q-g_1\tilde{r}
\end{eqnarray}
We neglect all additive constants to $w$ because they can be
absorbed in the rescaling of $x_0$ and $x_1$. The solution for $q$
approaches the fixed point I and II as $\tilde{r}\rightarrow \mp
\infty$, respectively.
\\
\indent At the fixed points, the six dimensional metric is given by
\begin{equation}
ds^2=e^{-2q_0-2g_1\tilde{r}}dx^2_{1,1}+e^{2q_0}d\tilde{r}^2+\frac{e^{2q_0}}{4g_1^2}\delta_{\alpha\beta}\sigma^\alpha\sigma^\beta\label{6dmetric}
\end{equation}
where $q_0$ is the value of $q$ at the fixed points. By rescaling
the $x^\mu$ and $\tilde{r}$ by a factor of $e^{-q_0}$ and
$-e^{q_0}$, respectively, we can write \eqref{6dmetric} as
\begin{equation}
ds^2=e^{\frac{2\bar{r}}{L}}dx^2_{1,1}+d\bar{r}^2+\frac{R^2}{4}\delta_{\alpha\beta}\sigma^\alpha\sigma^\beta
\end{equation}
which is the $AdS_3\times S^3$ metric. The radii of $AdS_3$ and
$S^3$ are given by $L=\frac{e^{q_0}}{g_1}$ and
$R=\frac{e^{q_0}}{g_1}$, respectively. The central charge in the
dual CFT is given by \cite{Brown Henneaux}
\begin{displaymath}
c=\frac{3L}{2G^{(3)}_N}\sim e^{4q_0}
\end{displaymath}
where we have used the relation between Newton constants in three
and six dimensions $G_N^{(3)}=\frac{G_N^{(6)}}{\textrm{Vol}(S^3)}$.
We find the ratio of the central charges
\begin{equation}
\frac{c_{{\rm{I}}}}{c_{{\rm{II}}}}=\frac{e^{4q_0}|_{{\rm{I}}}}{e^{4q_0}|_{\rm{II}}}=\frac{1}{1-\frac{g_1^2}{g_2^2}}>1.\label{YMCratio}
\end{equation}
From this equation, we find that the flow respects the c-theorem as
it should, and point I is the UV point while point II is the IR
point. Note that $d\bar{r}=-e^qd\tilde{r}$, so the UV and IR points
correspond to $\bar{r}\rightarrow \pm \infty$. We can interpret
$\bar{r}$ as an RG scale in the dual two dimensional field theory.
From the solutions for $q$, $\theta$ and $A$, we can check that the
relation \eqref{metric mathing} is satisfied. So, the solution is
indeed a solution of the theory considered in \cite{KK} and can be
obtained from three dimensional gauged supergravity. We also give
this solution in the three dimensional framework in appendix
\ref{3Dflow}.
\\ \indent We briefly look at the behavior of the scalar fields near
the UV point I. From \eqref{Asolfinall} and \eqref{gsolfinal}, we
find that
\begin{equation}
A\sim e^{2g_1\tilde{r}}\sim e^{-\frac{2\bar{r}}{L}} \,\,\,
\textrm{and}\,\,\, e^{q}\sim e^{-\frac{2\bar{r}}{L}}\, .
\end{equation}
We can see that the flow is driven by a vacuum expectation value of
a marginal operator of dimension two. Although this is not expected,
we will confirm this fact in appendix \ref{3Dflow} in which we will
reobtain this solution in the three dimensional gauged supergravity.
So, this flow is a vev. flow driven by a vacuum expectation value of
a marginal operator. Notice that gauge-field background of
\eqref{ansatz} that we have found corresponds to a single $SU(2)$
instanton on the four-space $(r,S^3)$, interpolating between winding
number 0 for $\bar{r}\rightarrow -\infty$ and winding number 1 for
$\bar{r}\rightarrow +\infty$. In the next section we will generalize
this result to a multi-instanton configuration for semisimple $G$
gauge fields, which therefore will not admit a three dimensional
interpretation. \\ \indent Throughout this section, we have mainly
studied the flow solution in the context of the $SU(2)$ reduction to
three dimensions. This leads to the form of the solution given
above. Before discussing the multi-instanton case, we would like to
change the form of the solution to make contact with what we will
find in the next section. First of all, we can change the
coordinates in \eqref{ansatz} to $R$ given by
\begin{equation}
\frac{dR}{dr}=-g_1Re^{-q}\, .
\end{equation}
We have put a minus sign in order to identify the UV point with
$R\rightarrow \infty$ and the IR with $R\rightarrow 0$. We then find
that the metric is given by
\begin{equation}
ds^2=e^{2w}(-dx_0^2+dx^2_1) +\frac{e^{2q}}{g_1R^2}dy^idy^i
\end{equation}
where $dy^idy^i=dR^2+\frac{R^2}{4}\sigma^\alpha\sigma^\alpha$ is the
flat metric of the four dimensional space. This is the form of the
metric we will see in the next section in which the 4-dimensional
part is conformally flat. The second point is the solution for $A$
in \eqref{Asol}. Recall that the relation between $R$ and
$\tilde{r}$ is $\frac{dR}{d\tilde{r}}=-g_1R$, we can write
\begin{equation}
A=\frac{\lambda^2}{g_2R(\lambda^2+R^2)}
\end{equation}
where we have chosen $C_2=\frac{1}{g_1}\ln
\big(-\frac{g_2}{\lambda^2}\big)$. This is a single instanton
solution at the origin $R=0$ in the polar coordinates. Notice that
this is the instanton solution in the singular gauge in which the
winding number come from the contribution near $R=0$. In the next
section, we will study a flow solution with $N$ instantons but in
the regular gauge.
%%%%%%%%%%%%%%%%%%%%%%%%%%%%%%%%%%%%%%%%%%%%%%%%%%%%%%%%%%%%%%%%%%%%%%%%%%%%%%%%%%%%%%%%%%%%%%%%%%%%%%%%%%%%%%%%%%%%%%%%%%%%%%%%%%%%%%%%%%%%%%%%%%%%%%%
\section{RG flow solutions and multi-instantons}\label{instanton}
In this section, we generalize the solution obtained in the previous
section by considering the gauge field configuration describing $N$
instantons. We will further make an extension to gauge fields of an
arbitrary gauge group $G$. The solution we will study is very
similar to the solution given in \cite{duff} and further studied in
\cite{Pope_instanton_moduli}. In this paper, we give an
interpretation of this solution in the context of an RG flow in the
dual two dimensional field theory. We start by reobtaining this
solution and then discuss its implication in term of the RG flow.
\subsection{Flow solutions}
Since we are going to use the full six-dimensional theory, we will
now turn on both $v^z$ and $\tilde{v}^z$ . Throughout this section,
we also assume that both $v^z$ and $\tilde{v}^z$ are positive. If
this is not the case, the phase transition discussed in \cite{duff}
is unavoidable. The supersymmetry transformations of fermions are
the same to leading order in fermionic fields and given by
\eqref{deltapsi}, \eqref{deltalambda} and \eqref{deltachi}. On the
other hand, the bosonic field equations are
\begin{eqnarray}
\hat{D}(e^{2\hat{\theta}}\hat{*}\hat{G}_3)+\tilde{v}^z \hat{F}^I\wedge \hat{F}^I&=&0,\label{G3eq1}\\
\hat{D}[(v^ze^{\hat{\theta}}+\tilde{v}^ze^{-\hat{\theta}})
\hat{*}\hat{F}^I]-2v^ze^{2\hat{\theta}}\hat{*}\hat{G}_3\wedge
\hat{F}^I+2\tilde{v}^z\hat{*}\hat{G}_3\wedge
\hat{F}^I&=&0,\label{FIeq1} \\
\hat{d}\hat{*}\hat{d}\hat{\theta}+(v^ze^{\hat{\theta}}+\tilde{v}^ze^{-\hat{\theta}})
\hat{*}\hat{F}^I\wedge
\hat{F}^I+2e^{2\hat{\theta}}\hat{*}\hat{G}_3\wedge\hat{G}_3&=&0\label{THeq1}
\, .
\end{eqnarray}
It is easy to see that if we set $v^z=1$, $\tilde{v}^z=0$ and take a
spherically symmetric single instanton configuration (i.e. an
instanton at the origin of $\mathbb{R}^4$) for the gauge field
$A^I$, then the above equations reduce to the ones discussed in the
previous section. The Bianchi identity is
\begin{equation}
\hat{D}\hat{G}_3=v^z\hat{F}^I\wedge \hat{F}^I. \label{Bianchi}
\end{equation}
We take an ansatz for the metric as
\begin{equation}
ds^2_6=e^{2f}(-dx^2_0+dx_1^2)+ds^2_4 \label{metric1}
\end{equation}
where $f$ only depends on the coordinates $z^\alpha$,
$\alpha=2,\ldots 5$ of the four dimensional metric
$ds^2_4=g_{\alpha\beta}dz^\alpha dz^\beta$. We first look at the
$\delta\lambda^I=0$ equation. We can satisfy this condition by
choosing $F^I_{\alpha\beta}$ to be self dual because of the
anti-self duality of the $\Gamma_{\alpha\beta}$, $\alpha,
\beta=2,\ldots 5$. The anti-self duality of $\Gamma_{\alpha\beta}$
is implied by the condition $\Gamma_7\epsilon=\epsilon$ and the
two-dimensional chirality $\Gamma_{01}\epsilon=\epsilon$ chosen in
$\delta\psi_\mu=0$ below. The indices $I,J\ldots=1,2,\ldots,
\textrm{dim}\, G$ are now $G$ adjoint indices. The gauge fields and
three form field strength are
\begin{eqnarray}
\hat{A}^I&=&A^I , \qquad \hat{F}^I=F^I, \nonumber \\
\hat{G}_3&=&G+dx_0\wedge dx_1\wedge d\Lambda \, .
\end{eqnarray}
The hatted fields are six dimensional ones while the unhatted fields
represented by differential forms without indices have only
components along $ds_4^2$. The $x_0$ and $x_1$ components will be
shown explicitly. The three form field satisfies the Bianchi
identity \eqref{Bianchi} which gives $D\, G=v^zF^I\wedge F^I$. The
dual of $\hat{G}_3$ is
\begin{equation}
\hat{*}\hat{G}_3=e^{-2f}*d\Lambda -e^{2f}dx_0\wedge dx_1\wedge *G
\end{equation}
where $\hat{*}$ and $*$ are Hodge duals in six and four dimensions,
respectively. We have used the same convention as \cite{nishino}
namely $\epsilon^{012345}=1$. Using equation \eqref{G3eq1}, we find
\begin{eqnarray}
D(e^{2\theta-2f}*d\Lambda)&=&v^zF^I\wedge F^I,\label{stardLambdaeq}\\
D(e^{2\theta+2f}*G)&=&0\Rightarrow
*G=e^{-2\theta-2f}d\tilde{\Lambda}\, .\label{starG}
\end{eqnarray}
We take $F^I$ to be self dual with respect to the four dimensional
$*$. This corresponds to an instanton configuration. The dual of
$\hat{F}^I$ is given by
\begin{equation}
\hat{*}\hat{F}^I=-e^{2f}dx_0\wedge dx_1\wedge *F^I\, .
\end{equation}
We now come to supersymmetry transformations. Using our ansatz and
the results given above, we find the Killing spinor equations
\begin{eqnarray}
\delta{\chi}&=&\frac{1}{2}\pd\Slash\, \theta\epsilon
-\frac{1}{12}e^\theta \hat{G}\Slash_3 \epsilon \nonumber \\
&=& \frac{1}{2}\pd\Slash\, \theta+\frac{1}{2}e^{\theta-2f}\pd\Slash
\Lambda -\frac{1}{2}e^{-\theta-2f}\pd \Slash \tilde{\Lambda} =0
\label{deltaChi} \\
\delta \psi_\mu &=&D_\mu \epsilon
+\frac{1}{24}e^{\theta}\hat{G}\Slash_3 \Gamma_\mu\epsilon,\, \mu=0,1
\nonumber
\\  &=& \frac{1}{2}\Gamma_\mu \pd \Slash \,
f-\frac{1}{4}e^{\theta-2f}\Gamma_\mu \pd \Slash \, \Lambda
-\frac{1}{4}e^{-\theta-2f}\Gamma_\mu\pd \Slash \tilde{\Lambda}=0\, .
\label{deltaPsiMu}
\end{eqnarray}
We have used the notation $\hat{G}\Slash_3\equiv
\Gamma^{MNP}\hat{G}_{3MNP}$ and a projector
$\Gamma_{2345}\epsilon=\epsilon$ which is also equivalent to
$\Gamma_{01}\epsilon=\epsilon$. This implies that the solution
preserves half of the six dimensional supersymmetry. Taking the
combination $[\eqref{deltaChi}-2\eqref{deltaPsiMu}]$, we find
\begin{equation}
2d\Lambda =-e^{-\theta+2f}d(\theta-2f).
\end{equation}
The solution is easily found to be
\begin{equation}
\Lambda=\frac{1}{2}e^{-\theta+2f}+C_1\label{Lambda}
\end{equation}
with a constant $C_1$. Similarly, the combination
$[\eqref{deltaChi}+2\eqref{deltaPsiMu}]$ gives
\begin{equation}
de^{\theta+2f}-2d\tilde{\Lambda}=0 \Rightarrow
\tilde{\Lambda}=\frac{1}{2}e^{\theta+2f}+C_2
\end{equation}
with a constant $C_2$. The equation from $\delta \psi_\alpha=0$
gives
\begin{equation}
D_\alpha\epsilon-\frac{1}{2}\pd\Slash \, f\Gamma_\alpha \epsilon=0\,
.\label{internalGravitino}
\end{equation}
We now make the following ansatz for the 4-dimensional metric
\begin{equation}
ds_4^2=e^{2g}dy^idy^i. \label{metric2}
\end{equation}
With the supersymmetry transformation parameter of the form
$\epsilon=e^{\frac{f}{2}}\tilde{\epsilon}$, we can write equation
\eqref{internalGravitino} as
\begin{equation}
\pd_i\tilde{\epsilon}-\frac{1}{2}\Gamma_{ji}\pd^j(f+g)\tilde{\epsilon}=0\,
.
\end{equation}
To satisfy this equation, we simply choose $g=-f$ and find that
$\tilde{\epsilon}$ is a constant spinor. So, we have solved all the
Killing spinor equations.
\\ \indent We can easily check that equation \eqref{FIeq1} is
identically satisfied with our explicit forms of $\Lambda$ and
$\tilde{\Lambda}$. We now solve equations \eqref{stardLambdaeq} and
the Bianchi identity \eqref{Bianchi}. We start with the $SU(2)$
instanton configuration from \cite{jackiw}
\begin{equation}
A_i=\frac{i}{2}\bar{\sigma}_{ij}\pd_j \ln \rho,\qquad \rho
=1+\sum_{a=1}^n\frac{\lambda_a^2}{(y-y_a)^2}\, .
\end{equation}
Notice that, we have rescaled the $A_i$ form \cite{jackiw} by a
factor of $\frac{1}{2}$. This can be done without loosing
generalities, see for example, \cite{lecture_instanton} for a
discussion. The $\bar{\sigma}_{ij}$ matrices are anti-self dual and
can be written in terms of Pauli matrices $\sigma_x$ as
\cite{jackiw}
\begin{equation}
\bar{\sigma}_{xy}=\frac{1}{4i}[\sigma_x,\sigma_y], \qquad
\bar{\sigma}_{x4}=-\frac{1}{2}\sigma_x,\,\,\, x,y=1,2,3\, .
\end{equation}
This solution can be generalized to any semi-simple group $G$ to
obtain the solution given in \cite{SUN_instanton}. We can write the
above $\bar{\sigma}_{ij}=\bar{\eta}^I_{ij}t^I$ where
$\bar{\eta}^I_{ij}$ and $t^I$ are 't Hooft's tensor generating
$SU(2)$ subgroup of $SO(4)$ and $SU(2)$ generators, respectively.
For any group $G$, the solution is \cite{SUN_instanton}
\begin{equation}
A^I_i=G^I_a\bar{\eta}^a_{ij}\pd_j\ln \rho,\,\,\, a=1,2,3 \, .
\end{equation}
For $G=SU(2)$, we simply have $G^I_a=\delta^I_a$. The solution for
any group $G$ can be obtained by embedding $SU(2)$ in $G$
\cite{SUN_instanton}. In order to solve \eqref{stardLambdaeq} and
\eqref{Bianchi}, we need to compute $F^I_{ij}F^{Iij}$. For $SU(2)$
instanton, we have \cite{jackiw}
\begin{equation}
F^I_{ij}F^{Iij}=-\square\square \ln\rho\, .
\end{equation}
For $G$ instanton, the result is the same up to some numerical
factors, from \cite{SUN_instanton},
\begin{equation}
F^I_{ij}F^{Iij}=-\frac{2}{3}c(G)d(G)\square\square \ln\rho\, .
\end{equation}
\\ \indent
We now come to the Bianchi identity for $G$, $DG=v^zF^I\wedge F^I$,
which gives
\begin{equation}
\square
e^{-(\theta+2f)}=-v^zF^I_{ij}F^{Iij}=v^z\frac{2}{3}c(G)d(G)\square
\square \ln \rho \label{eq1}
\end{equation}
where we have used \cite{jackiw}
\begin{equation}
*(F^I\wedge F^I)=*(*F^I\wedge
F^I)=\frac{1}{2}F^I_{ij}F^{Iij}=-\frac{1}{3}c(G)d(G)\square\square
\ln\rho\, .
\end{equation}
We can solve \eqref{eq1} and obtain
\begin{eqnarray}
e^{-(\theta+2f)}&=&\frac{d}{r^2}+v^z\bigg(\frac{2}{3}c(G)d(G)\square
\ln \rho+\sum_{a=1}^n\frac{4}{(y-y_a)^2}\bigg)\nonumber
\\ &\equiv&\frac{d}{r^2}+v^z\frac{2}{3}c(G)d(G)\square\ln \tilde{\rho}\,
.\label{t+2f}
\end{eqnarray}
We have removed the singularities in the solution by defining
$\tilde{\rho}=\rho\prod_{a=1}^n(y-y_a)^2$. Inserting $\Lambda$ from
\eqref{Lambda} in \eqref{stardLambdaeq}, with $*d\Lambda$ replaced
by $e^{-2f}*d\Lambda$, gives
\begin{equation}
\square
e^{\theta-2f}=-\tilde{v}^zF^I_{ij}F^{Iij}=\tilde{v}^z\frac{2}{3}c(G)d(G)\square\square\ln
\rho\, .
\end{equation}
The solution is similar to the previous equation
\begin{equation}
e^{\theta-2f}=\frac{c}{r^2}+\tilde{v}^z\frac{2}{3}c(G)d(G)\square\ln\tilde{\rho}\label{t-2f}
\end{equation}
where $c$ is an integration constant. The two integration constants
$c$ and $d$ are proportional, respectively,  to the fluxes of
$\hat{*}\hat{G}_3$ and $\hat{G}_3$ through the $S^3$. Therefore,
they represent, respectively, the number of $D1$ and $D5$ branes. We
can directly see this by considering for example, the $\hat{G}_3$
flux near $r=0$
\begin{eqnarray}
Q_1&=&
\frac{1}{8\pi^2}\int_{S^3}e^{2\theta}\hat{*}\hat{G}_3= \frac{c}{4} \\
Q_5&=& \frac{1}{8\pi^2}\int_{S^3}\hat{G}_3 =\frac{d}{4}\, .
\end{eqnarray}
We have used the same normalization of $Q_1$ and $Q_5$ as in
\cite{Pope_instanton_moduli}. Indeed, we can regard our six
dimensional theory as a subsector of type I theory compactified on
$K3$. In this solution, we have $D5$ branes wrapped on $K3$ and $D1$
branes transverse to it. The solution we give here is the same as
the gauge dyonic string studied in \cite{duff} and
\cite{Pope_instanton_moduli}.
\\ \indent
The behaviors of $e^{-4f}$ near $r\rightarrow \infty$ and
$r\rightarrow 0$ are given by
\begin{eqnarray}
r\rightarrow \infty: & &e^{-4f}\sim \frac{(c+4\tilde{v}^zN)(d+4v^zN)}{r^4}=\frac{16(Q_1+\tilde{v}^zN)(Q_5+v^zN)}{r^4},\label{fluctuation} \\
r\rightarrow 0: & &e^{-4f}\sim \frac{cd}{r^4}=\frac{16Q_1Q_5}{r^4}\,
.
\end{eqnarray}
We have introduced the instanton number $N$ given by
\begin{equation}
N=\frac{1}{32\pi^2}\int
d^4y(*F)_{ij}F^{ij}=-\frac{1}{48\pi^2}c(G)d(G)\int d^4y
\square\square \ln \tilde{\rho} \, .
\end{equation}
At the fixed points, the metric is given by
\begin{equation}
ds^2_6=\frac{r^2}{L^2}(-dx_0^2+dx_1^2)+\frac{L^2}{r^2}dr^2+L^2ds^2(S^3).\label{6dmetricAtfixpoints}
\end{equation}
where we have rewritten the four dimensional flat metric in the
polar coordinates
\begin{equation}
dy^idy^i=dr^2+r^2ds^2(S^3).
\end{equation}
The metric \eqref{6dmetricAtfixpoints} is readily seen to be
$AdS_3\times S^3$ metric with the $AdS_3$ and $S^3$ having the same
radius $L$. The AdS radii at the fixed points near $r\sim\infty$ and
$r\sim0$ are $L=[(c+4\tilde{v}^zN)(d+4v^zN)]^{\frac{1}{4}}$ and
$L=(cd)^{\frac{1}{4}}$, respectively. In the dual two dimensional
conformal field theory, this solutions describes an RG flow from the
CFT UV to the CFT IR with the ratio of the central charges
\begin{equation}
\frac{c|_0}{c|_\infty}=\frac{e^{-4f}|_0}{e^{-4f}|_\infty}=\frac{cd}{(c+4\tilde{v}^zN)(d+4v^zN)}<1
\label{cc}
\end{equation}
where we have used the relation between the central charge and AdS
radius $c\sim
\frac{L}{G_N^{(3)}}\sim\frac{L\textrm{Vol}(S^3)}{G_N^{(6)}}\sim
e^{-4f}$. The UV point corresponds to $r=\infty$ while the IR point
is at $r=0$.
\\ \indent To extract the dimension of the operator
driving the flow, we need to consider the behavior of the
fluctuation of the metric around $AdS_3\times S^3$ near the UV
point. To simplify the manipulation, we first consider here a single
instanton at the origin $y_a^i=0$. With this simplification, the
solution, up to group theory factors which are not relevant for this
discussion, is given by
\begin{equation}
e^{-4f}=\bigg(\frac{c}{r^2}+\tilde{v}^z\square \ln
(r^2+\lambda^2)\bigg)\bigg(\frac{d}{r^2}+v^z\square \ln
(r^2+\lambda^2)\bigg).
\end{equation}
As $r\rightarrow \infty$, the solution behaves
\begin{eqnarray}
e^{-4f}&\sim&
\frac{(c+4\tilde{v}^z)(d+4v^z)}{r^4}\bigg(1+\frac{8\lambda^2[c+d+4(\tilde{v}^z+v^z)]}{r^2(c+4\tilde{v}^z)(d+4v^z)}+\ldots\bigg)\nonumber \\
\textrm{or}\qquad e^{2g}&=&e^{-2f}\nonumber \\ &\sim &
\frac{\sqrt{(c+4\tilde{v}^z)(d+4v^z)}}{r^2}\bigg(1+\frac{4\lambda^2[c+d+4(\tilde{v}^z+v^z)]}{r^2(c+4\tilde{v}^z)(d+4v^z)}+\ldots\bigg).
\end{eqnarray}
From this equation, we find the fluctuation
\begin{equation}
\delta e^g\sim
\frac{2\lambda^2[c+d+4(\tilde{v}^z+v^z)]}{r^2(c+4\tilde{v}^z)(d+4v^z)}
\end{equation}
which give $\Delta=2$ in agreement with the result of the previous
section. We can also see this in the coordinate
$\tilde{r}=[(c+4\tilde{v}^z)(d+4v^z)]^{\frac{1}{4}}\ln r$ in which
\begin{equation}
ds^2_6=e^{\frac{2\tilde{r}}{L}}(-dx_0^2+dx_1^2)+d\tilde{r}^2+R^2ds(S^3)^2
\end{equation}
and $\delta e^g\sim e^{\frac{-2\tilde{r}}{L}}$. We have identified
the $AdS_3$ and $S^3$ radii
$L=R=[(c+4\tilde{v}^z)(d+4v^z)]^{\frac{1}{4}}$. In the general case
with $N$ instantons, it can be checked through a more complicated
algebra that the fluctuation of the metric behaves as $\sim r^{-2}$
near the UV point. This can be seen as follows. $\tilde{\rho}$ have
an expansion in powers of $r^{2n}+r^{2n-2}$ with $r^2=y_iy_i$. From
this, we find that $\square\ln \tilde{\rho}\sim
\frac{1}{r^{4n}}(r^2+r^4+\ldots r^{4n-2})$ from which we see that
$r^{-2}$ is the leading term we have found in \eqref{fluctuation}
while the subleading $r^{-4}$ gives $\Delta=2$ as in the single
instanton case. So, our flow is a vev. flow driven by a vacuum
expectation value of a marginal operator. \\ \indent We end this
subsection by giving a comment on the anti-instanton gauge field
configuration. We need to choose the three dimensional chirality
$\Gamma_{01}\epsilon=-\epsilon$ which implies the self dual
$\Gamma_{\alpha\beta}$ from
$\Gamma_7\epsilon=-\Gamma_{2345}\epsilon=\epsilon$. So, the
condition $\delta \lambda^I=0$ is still satisfied. The BPS equations
\eqref{deltaChi} and \eqref{deltaPsiMu} are modified by some sign
changes. We find the following equations
\begin{eqnarray}
\frac{1}{2}\pd\Slash\, \theta-\frac{1}{2}e^{\theta-2f}\pd\Slash
\Lambda +\frac{1}{2}e^{-\theta-2f}\pd \Slash \tilde{\Lambda} &=&0 \\
\frac{1}{2} \pd \Slash \, f+\frac{1}{4}e^{\theta-2f} \pd \Slash \,
\Lambda +\frac{1}{4}e^{-\theta-2f}\pd \Slash \tilde{\Lambda}&=&0\, .
\end{eqnarray}
This change results in an extra minus sign in
$\Lambda=-\frac{1}{2}e^{-\theta+2f}+C_1$. The field strength
$*F^I=-F^I$ gives an extra minus sign in equation $DG=v^zF^I\wedge
F^I$. The final result is
\begin{eqnarray}
e^{-(\theta+2f)}&=&\frac{d}{r^2}-v^z\frac{2}{3}c(G)d(G)\square
\ln \tilde{\rho}\\
e^{\theta-2f}&=&\frac{c}{r^2}-\tilde{v}^z\frac{2}{3}c(G)d(G)\square\ln\tilde{\rho}
\end{eqnarray}
with the behavior near the fixed points
\begin{eqnarray}
r\rightarrow \infty: & &e^{-4f}\sim \frac{(c-4\tilde{v}^zN)(d-4v^zN)}{r^4}, \\
r\rightarrow 0: & &e^{-4f}\sim \frac{cd}{r^4}\, .
\end{eqnarray}
In this case, $N$ is now negative.
%%%%%%%%%%%%%%%%%%%%%%%%%%%%%%%%%%%%%%%%%%%%%%%%%%%%%%%%%%%%%%%%%%%%%%%%%%%%%%%%%%%%%%%%%%%%%%%%%%%%%%%%%%%%%%%%%%%%%%%%%%%%%%%%%%%%%%%%%%
\subsection{Central charges of the dual CFT}\label{CentralCharge}
We now give some comments on the central charge of the dual (4,0)
CFT. We have mentioned that solutions to the six-dimensional
supergravity given in the previous sections can be interpreted as a
D1/D5 brane system in type I string theory on K3. As type I and
heterotic theories are S-dual to each other \cite{polchinski and
witten}, this D1/D5 system is dual to the F1/NS5 brane system in the
heterotic theory. We choose to work with heterotic string theory on
K3 with the string frame effective action given by \cite{witten
duff}
\begin{eqnarray}
I_6&=&\frac{(2\pi)^3}{\alpha'^2}\int
d^6x\sqrt{-g}e^{2\theta}\big[R_6+4\pd_M\theta
\pd^M\theta-\frac{1}{12}G_{MNP}G^{MNP}\big]\nonumber \\
& &+\int_{M_6}\frac{1}{4(2\pi)^3\alpha'}B\wedge
\sum_\alpha\tilde{v}^\alpha\textrm{tr}F_\alpha\wedge F_\alpha
\label{6Daction}
\end{eqnarray}
where we have given only the relevant terms for our discussion. All
the notations are the same as those in \cite{witten duff} including
the modified three-form field strength
\begin{equation}
G=dB-\frac{\alpha'}{4}\sum_\alpha v^\alpha \Omega(F^\alpha)
\end{equation}
where $\Omega(F^\alpha)$ is the Chern-Simons term of the gauge field
$A^\alpha$. \\ \indent To compute the central charge, we need to
know the coefficient of the Einstein-Hilbert term. The central
charge is then given by \cite{Brown Henneaux}
\begin{equation}
c=\frac{3\ell}{2G_N^{(3)}}\label{central charge}
\end{equation}
where $\ell$ is the $AdS_3$ radius. Note that the central charge can
be written as $c=24\pi \alpha$ with $\alpha$ being the coefficient
of the Einstein-Hilbert term. In appendix \ref{derivation}, we give
a derivation of this result by using the computation of chiral
correlators involving $T_{zz}$ (${\bar T}_{{\bar z}{\bar z}}$) in
the spirit of \cite{Witten}.
\\ \indent Our ansatz for
the metric is
\begin{equation}
ds^2_6=\ell^2
\bigg(\frac{d\rho^2}{4\rho^2}+\frac{1}{\rho}dx^2_{1,1}\bigg)+\ell^2ds^2(S^3).
\end{equation}
We have used the $AdS_3$ metric in the form given in
\cite{skenderis}. The three-form field is
\begin{equation}
G=\ell^2 h \epsilon_3+\ell^2a\omega_3
\end{equation}
where $\epsilon_3$ and $\omega_3$ are the volume form on the unit
$AdS_3$ and $S^3$, respectively. The radii of $AdS_3$ and $S^3$ are
the same by the equation of motion for $\theta$ with constant
$\theta$. The $G$ equations of motion and Bianchi identity require
$h$ and $a$ to be constant. Einstein equation determines the value
of $h=a=2$. The fluxes of $G$ and $*G$ are given by
\begin{eqnarray}
Q_1&=&\frac{(2\pi)^3}{\alpha'^2}\int_{S^3}e^{2\theta}*G=\frac{(2\pi)^5\ell^2e^{2\theta}}{\alpha'^2},
\\
Q_5&=&\frac{(2\pi)^3}{\alpha'^2}\int_{S^3}G=\frac{(2\pi)^5\ell^2}{\alpha'^2}\,
.
\end{eqnarray}
In order to relate $Q_1$ and $Q_5$ to the number of F1 strings and
NS5 branes, $N_1$ and $N_5$, we match our ansatz to the F1/NS5
solution in six dimensions given in \cite{F1NS5}. The three-form
field strength is
\begin{equation}
G=2\alpha' N_5 (\textrm{Vol}_{AdS_3}+\textrm{Vol}_{S^3})\, .
\end{equation}
We find that, after matching the flux of this solution with that of
our ansatz,
\begin{equation}
N_1=2\pi\alpha' Q_1,\qquad N_5=\frac{\alpha' Q_5}{(2\pi)^5}\, .
\end{equation}
\\ \indent
We now make a reduction of \eqref{6Daction} on $S^3$ and obtain
\begin{eqnarray}
I_3&=&\frac{\alpha'^2Q_1Q_5}{2(2\pi)^5}\int d^3x\sqrt{g}
R+\frac{\alpha'Q_1}{4}\int_{M3}v\Omega(F^\alpha)
+\frac{\alpha'Q_5}{4(2\pi)^6}\int_{M3}\tilde{v}\Omega(F^\alpha) \nonumber \\
& & -\frac{\alpha'Q_1Q_5}{8(2\pi)^5}\int_{M3}\Omega(F^I) \nonumber
\\  &=& \frac{N_1N_5}{4\pi}\int d^3x\sqrt{g}
R+\frac{N_1}{8\pi}\int_{M3}v\Omega(F^\alpha)+\frac{N_5}{8\pi}\int_{M3}\tilde{v}\Omega(F^\alpha)
\nonumber \\
& &-\frac{N_1N_5}{2(8\pi)}\int_{M3}\Omega (F^I)\label{3Daction}
\end{eqnarray}
where we have given only the Einstein-Hilbert and Chern-Simons terms
which are relevant for the present discussion. The $SU(2)$
Chern-Simons term $\Omega(F^I)$ cannot be determined by the
dimensional reduction of the action \eqref{6Daction}. As in our
previous work \cite{KK}, its presence in the effective action is
implied by the equation of motion for $F^I$. \\ \indent We can now
use \eqref{3Daction}, together with the results in \cite{skenderis0,kraus},
which we will rederive in a different way in the Appendix
\ref{derivation},  to compute the central charges
\begin{equation}
c_L=6N_1N_5, \qquad c_R=6N_1N_5.
\end{equation}
The Kac-Moody levels of the $SU(2)$ and gauge group $G_\alpha$ can
be computed from the Chern-Simons terms of the $SU(2)$ and
$G^\alpha$ gauge fields. Using the result in appendix
\ref{derivation}, we find
\begin{eqnarray}
\textrm{SU(2) level}&:& k_{SU(2)}=8\pi \beta=N_1N_5, \\
\textrm{G$_\alpha$ level}&:& k_\alpha= 2(v^\alpha
N_1+\tilde{v}^\alpha N_5).
\end{eqnarray}
%%%%%%%%%%%%%%%%%%%%%%%%%%%%%%%%%%%%%%%%%%%%%%%%%%%%%%%%%%%%%%%%%%%%%%%%%%%%%%%%%%%%%%%%%%%%%%%%%%%%%%%%%%%%%%%%%%%%%%%%%%%%%%%%%%%%%%%%%%
\section{Conclusions}\label{conclusion}
In this paper, we have found three analytic RG flow solutions in six
and three dimensional supergravities. In six dimensional
supergravity, we have found the solution in which the internal
components, outside the $AdS_3$ part, of the gauge fields, describe
a configuration with $N$ instantons. We have discussed separately
the case $N=1$. This is interesting in the sense that the solution
can be obtained from uplifting the three dimensional solution. We
have also given the corresponding solution in the Chern-Simons
gauged supergravity. With the reduction given in \cite{KK}, the
solution can be lifted to six dimensions and easily
seen that it is indeed the same as the six dimensional solution. \\
\indent The flows describe a deformation of the UV CFT by a vacuum
expectation value of a (not exactly) marginal operator.
Interestingly, these RG flows have an interpretation in terms of
Yang-Mills instantons tunnelling between $|N\rangle$ Yang-Mills
vacuum in the UV and $|0\rangle$ in the IR, and this fact is in turn
related to the different values of the central charge at the two
fixed points. In the general $N$ instanton solution, there is a
subtlety of phase transitions occurring whenever $v$ and $\tilde{v}$
change sign. We have avoided this issue by assuming the positivity
of both $v$ and $\tilde{v}$. We do not have a clear interpretation
of this phase transition in the dual CFT, so it would be interesting
to study this issue in more detail. We finally give some comments on
the central charge of the dual CFT along with the derivation of the
central charge of the boundary CFT from the gravity theory in the
bulk.

\acknowledgments This work has been supported in part by the EU
grant UNILHC-Grant Agreement PITN-GA-2009-237920.
%%%%%%%%%%%%%%%%%%%%%%%%%%%%%%%%%%%%%%%%%%%%%%%%%%%%%%%%%%%%%%%%%%%%%%%%%%%%%%%%%%%%%%%%%%%%%%%%%%%%%%%%%%%%%%%%%%%%%%%%%%%%%%%%
\appendix
\section{Flow solution from $N=4$ three dimensional gauged
supergravity}\label{3Dflow} In this appendix, we study a flow
solution in $N=4$ three dimensional $(SO(3)\ltimes
\mathbf{R}^3)\times (G\ltimes \mathbf{R}^{\textrm{dim}G})$
Chern-Simons gauged supergravity. It has been shown in \cite{KK}
that this theory is equivalent to $SO(3)\times G$ Yang-Mills gauged
theory obtained from $SU(2)$ reduction of six dimensional
supergravity whose flow solution has been studied in section
\ref{6Dflow}. We are interested in the case $G=SO(3)$. We will see
that the solution we are going to find is the same as that in
section \ref{6Dflow} but now in another framework. As in our
previous papers, we use the formulation of \cite{dewit}. See
\cite{nicolai2, nicolai3} for more details of various gaugings.
\subsection{$(SO(3)\ltimes \mathbf{R}^3)\times (SO(3)\ltimes \mathbf{R}^3)$ gauged supergravity}
We now construct three dimensional gauged supergravity with gauge
groups $(SO(3)\ltimes \mathbf{R}^3)\times (SO(3)\ltimes
\mathbf{R}^3)$. This can be obtained from the theory constructed in
\cite{KK} by setting $G=SO(3)$. The scalar fields are described by
$\frac{SO(4,4)}{SO(4)\times SO(4)}$ coset manifold. We parametrize
the coset by
\begin{equation}
L=\left(
    \begin{array}{cccc}
      \frac{1}{1-A(r)^2}\mathbf{I}_{3\times 3} & 0 & \frac{A(r)}{1-A(r)^2}\mathbf{I}_{3\times 3} & 0 \\
      0 & \cosh{h(r)} & 0 & \sinh{h(r)} \\
      \frac{A(r)}{1-A(r)^2}\mathbf{I}_{3\times 3} & 0 & \frac{1}{1-A(r)^2}\mathbf{I}_{3\times 3} & 0 \\
      0 & \sinh{h(r)} & 0 & \cosh{h(r)} \\
    \end{array}
  \right).
\end{equation}
The $SO(3)$ generators are given by
\begin{equation}
J_{a_1}^A=\epsilon_{ABC}e_{BC}, \qquad
J_{a_2}^A=\epsilon_{ABC}e_{B+4,C+4},\,\,\, A, B, C =1, 2, 3
\end{equation}
where $(e_{AB})_{CD}=\delta_{AC}\delta_{BD}$ are $8\times 8$
matrices. The translational symmetries are generated by
\begin{eqnarray}
J_{b_1}^A&=&-e_{A4}+e_{A8}+e_{4A}+e_{8A} \nonumber \\
J_{b_2}^A&=&-e_{4,4+A}+e_{A+4,4}-e_{A+4,8}+e_{8,A+4},\,\,\, A=1,2,3.
\end{eqnarray}
$SO(4)$ R-symmetry generators are given by
\begin{equation}
J_\pm^{IJ}=J^{IJ}\pm\frac{1}{2}\epsilon_{IJKL}J^{KL},\qquad
J^{IJ}=e_{IJ}-e_{JI},\,\,\, I,J,\ldots =1,\ldots 4.
\end{equation}
In the $N=4$ theory, the $SO(4)$ R-symmetry decomposes to $SO(3)_+$
and $SO(3)_-$, and each factors acts separately on the two scalar
target spaces. In our case called the degenerate case, there is only
one target space, so we have only one $SO(3)$ which we will denote
by $SO(3)_+$. Non compact generators of $SO(4,4)$ are
\begin{equation}
Y_{ab}=e_{a,b+4}+e_{b+4,a},\,\, a,b =1,2,3.
\end{equation}
We now proceed as in \cite{gkn} using the formulation of
\cite{dewit}. The embedding tensor gives the following T-tensors,
$T_{\mathcal{A}\mathcal{B}}=\Theta_{\mathcal{M}\mathcal{N}}\mathcal{V}^{\mathcal{M}}_{\phantom{a}\mathcal{A}}\mathcal{V}
^{\mathcal{N}}_{\phantom{a}\mathcal{B}}$,
\begin{eqnarray}
T^{IJ,KL}&=&g_1(\mathcal{V}_{\textrm{a}_1}^{AIJ}\mathcal{V}_{\textrm{b}_1}^{AKL}+\mathcal{V}_{\textrm{b}_1}^{AIJ}\mathcal{V}_{\textrm{a}_1}^{AKL})
+g_2(\mathcal{V}_{\textrm{a}_2}^{AIJ}\mathcal{V}_{\textrm{b}_2}^{AKL}+\mathcal{V}_{\textrm{b}_2}^{AIJ}\mathcal{V}_{\textrm{a}_2}^{AKL})\nonumber
\\ &
&+h_1
\mathcal{V}_{\textrm{b}_1}^{AIJ}\mathcal{V}_{\textrm{b}_1}^{AKL}+h_2
\mathcal{V}_{\textrm{b}_2}^{AIJ}\mathcal{V}_{\textrm{b}_2}^{AKL}, \\
T^{IJ}_{ab}&=&g_1(\mathcal{V}_{\textrm{a}_1}^{AIJ}{\mathcal{V}_{\textrm{b}_1}}^{A}_{ab}
+\mathcal{V}_{\textrm{b}_1}^{AIJ}{\mathcal{V}_{\textrm{a}_1}}^{A}_{ab})
+g_2(\mathcal{V}_{\textrm{a}_2}^{AIJ}{\mathcal{V}_{\textrm{b}_2}}^{A}_{ab}
+\mathcal{V}_{\textrm{b}_2}^{AIJ}{\mathcal{V}_{\textrm{a}_2}}^{A}_{ab})\nonumber
\\ &
&+h_1\mathcal{V}_{\textrm{b}_1}^{AIJ}{\mathcal{V}_{\textrm{b}_1}}^A_{ab}
+h_2\mathcal{V}_{\textrm{b}_2}^{AIJ}{\mathcal{V}_{\textrm{b}_2}}^A_{ab}.
\end{eqnarray}
With the coset representative $L$, we can compute all the needed
quantities using
\begin{eqnarray}
L^{-1} D_\mu L&=& \frac{1}{2}Q^{IJ}_\mu X^{IJ}+Q^\alpha_\mu
X^{\alpha}+e^A_\mu Y^A\, , \nonumber \\
L^{-1}t^\mathcal{M}L&=&\frac{1}{2}\mathcal{V}^{\mathcal{M}IJ}X^{IJ}+\mathcal{V}^\mathcal{M}_{\phantom{a}\alpha}X^\alpha+
\mathcal{V}^\mathcal{M}_{\phantom{a}A}Y^A\, .\label{coset}
\end{eqnarray}
The consistency condition from supersymmetry,
$\mathbb{P}_{\boxplus}T^{IJ,KL}=0$, \cite{dewit} requires that
$h_2=-h_1$. The resulting $\mathcal{V}^\mathcal{M}_\mathcal{A}$ are
given by
\begin{eqnarray}
\mathcal{V}_{
\textrm{a}_{1,2}}^{AKL}&=&-\frac{1}{2}\textrm{Tr}[L^{-1}J_{\textrm{a}_{1,2}}^ALJ_+^{KL}],\nonumber\\
\mathcal{V}_{
\textrm{b}_{1,2}}^{AKL}&=&-\frac{1}{2}\textrm{Tr}[L^{-1}J_{\textrm{b}_{1,2}}^ALJ_+^{KL}],\nonumber\\
\mathcal{V}_{\textrm{a}_{1,2}}^{Aab}&=&\frac{1}{2}\textrm{Tr}[L^{-1}J_{\textrm{a}_{1,2}}^AL(e_{a,b+4}+e_{b+4,a})],\nonumber\\
\mathcal{V}_{\textrm{b}_{1,2}}^{Aab}&=&\frac{1}{2}\textrm{Tr}[L^{-1}J_{\textrm{b}_{1,2}}^AL(e_{a,b+4}+e_{b+4,a})]
\end{eqnarray}
where $A, B, \ldots=1, 2, 3$ label $SO(3)$ gauge generators, and a
pair of indices $a,b, \ldots=1, 2, 3$ labels target space
coordinates. With an appropriate normalization, the tensor $f^{IJ}$
is given by
\begin{equation}
f^{IJ}_{ab,cd}=2\textrm{Tr}(e_{ba}J_+^{IJ}e_{cd}).
\end{equation}
With all these ingredients, we can now compute $A_1$ and $A_2$
tensors which give the scalar potential via
\begin{eqnarray}
A_1^{IJ}&=&-\frac{4}{N-2}T^{IM,JM}+\frac{2}{(N-1)(N-2)}\delta^{IJ}T^{MN,MN},\nonumber\\
A_{2j}^{IJ}&=&\frac{2}{N}T^{IJ}_{\phantom{as}j}+\frac{4}{N(N-2)}f^{M(I
m}_{\phantom{as}j}T^{J)M}_{\phantom{as}m}+\frac{2}{N(N-1)(N-2)}\delta^{IJ}f^{KL\phantom{a}m}_{\phantom{as}j}T^{KL}_{\phantom{as}m},\nonumber\\
V&=&\frac{4}{N}(A_1^{IJ}A_1^{IJ}-\frac{1}{2}Ng^{ij}A_{2i}^{IJ}A_{2j}^{IJ}).\label{a1a2v}
\end{eqnarray}
The potential for these two scalars is given by
\begin{equation}
V=e^{-4h}\bigg[h_1^2+\frac{2e^{2h}}{(A^2-1)^3}(g_1^2+A^2(g_2A(4g_1+g_2A(A^2-3))-3g_1^2))\bigg].\label{potential40}
\end{equation}
This simple looking potential admits five different critical points.
We can identify supersymmetric critical points by using the
procedures explained in \cite{gkn}. All non trivial critical points
are given in table I.\\   \textbf{Table I}: Critical points of the
potential \eqref{potential40}. $A_0$ and $h_0$ are vacuum
expectation values at the critical point of $A$ and $h$,
respectively.
\begin{center}
\begin{tabular}{|c|c|c|c|c|}
  \hline
  Critical points & $A_0$ & $h_0$ & $V_0$ & Preserved supersymmetries \\ \hline
  I & $0$ & $\ln{\big(-\frac{h_1}{g_1}\big)}$ & $-\frac{g_1^4}{h_1^2}$ & non supersymmetric \\ \hline
  II & $0$ & $\ln \big(\frac{h_1}{g_1}\big)$ & $-\frac{g_1^4}{h_1^2}$ & (4,0) \\ \hline
  III & $\frac{g_1}{g_2}$ & $\ln \bigg(\frac{\sqrt{-h_1^2(g_1^2-g_2^2)}}{g_1g_2}\bigg)$ & $-\frac{g_1^4g_2^4}{(g_1^2-g_2^2)^2h_1^2}$ & (4,0) \\ \hline
  IV & $\frac{g_1}{g_2}$ & $\ln \bigg(-\frac{\sqrt{-h_1^2(g_1^2-g_2^2)}}{g_1g_2}\bigg)$ & $-\frac{g_1^4g_2^4}{(g_1^2-g_2^2)^2h_1^2}$ & non supersymmetric \\ \hline
  V & $\frac{g_2}{g_1}$ & $\ln \frac{h_1\sqrt{g_1^2-g_2^2}}{\sqrt{g_1^4-g_1^2g_2^2+g_2^4}}$ & $-\frac{g_1^4-g_1^2g_2^2+g_2^4}{(g_1^2-g_2^2)^2h_1^2}$ & non supersymmetric \\ \hline
\end{tabular}
\end{center}
\indent Supersymmetric flow equations can be obtained from
supersymmetry transformations of fermions, $\delta\psi^I_\mu=0$ and
$\delta\chi^{iI}=0$. The metric ansatz is
\begin{equation}
ds^2=e^{2f}dx_{1,1}^2+dr^2\, .
\end{equation}
Recall that
\begin{eqnarray}
\delta\psi^I_\mu
&=&\mathcal{D}_\mu\epsilon^I+A_1^{IJ}\gamma_\mu\epsilon^J,\nonumber\\
\delta\chi^{iI}&=&
\frac{1}{2}(\delta^{IJ}\mathbf{1}-f^{IJ})^i_{\phantom{a}j}{\mathcal{D}{\!\!\!\!/}}\phi^j\epsilon^J
-NA_2^{JIi}\epsilon^J,\label{susyvar}
\end{eqnarray}
together with the formulae in equation \eqref{a1a2v}, we find, from
$\delta \chi^{Ii}$,
\begin{eqnarray}
\frac{dA}{dr}&=&\frac{2e^{-h}A(g_2A-g_1)}{\sqrt{1-A^2}}\label{Aeq}\\
\frac{dh}{dr}&=&\frac{2e^{-2h}}{(1-A^2)^{\frac{3}{2}}}[e^h(g_1-g_2A^3)+h_1(A^2-1)\sqrt{1-A^2}].\label{heq}
\end{eqnarray}
We can easily check that \eqref{Aeq} and \eqref{heq} have two
critical points which are exactly the same as II and III points in
table I. With a new function $g$ and new coordinate $\tilde{r}$
given by
\begin{equation}
g=h+\ln\sqrt{1-A^2}\qquad \textrm{and} \qquad
d\tilde{r}=e^{-g}dr,\label{define_g}
\end{equation}
we can write \eqref{Aeq} as
\begin{equation}
A'=\frac{dA}{d\tilde{r}}=2A(g_2A-g_1).
\end{equation}
The solution for $A$ is
\begin{equation}
A=-\frac{g_1}{e^{2g_1\tilde{r}+g_1C_1}-g_2}\, .
\end{equation}
As in section \ref{6Dflow}, we choose $C_1=\frac{1}{g_1}\ln{(-g_2)}$
and end up with
\begin{equation}
A=\frac{g_1}{(e^{2g_1\tilde{r}}+1)g_2}\, .\label{solA}
\end{equation}
With \eqref{define_g} and \eqref{solA}, equation \eqref{heq} can be
rewritten as
\begin{eqnarray}
g'=\frac{dg}{d\tilde{r}}&=&-2 \bigg[g_1 \bigg(-\frac{g_1 h_1
e^{-g}}{g_2^2 (e^{2 g_1 \tilde{r}}+1)^2}+\frac{g_1+g_2}{g_2 e^{2 g_1
\tilde{r}}+g_1+g_2}+\frac{1}{\frac{g_2 e^{2 g_1
\tilde{r}}}{g_2-g_1}+1}-2\bigg)\nonumber \\ & &+h_1 e^{-g}+g_1 \tanh
(g_1 \tilde{r})\bigg].
\end{eqnarray}
This can be solved, and the solution is
\begin{equation}
g=\ln
\bigg[\frac{(h_1+16C_2g_1e^{2g_1\tilde{r}})(g_2^2e^{4g_1\tilde{r}}+2g_2^2e^{2g_1\tilde{r}}-g_1^2+g_2^2)}{g_1g_2^2(1+e^{g_1\tilde{r}})^2}\bigg].
\end{equation}
This solution interpolates between II and III fixed points provided
that we choose $C_2=0$. We now move to $\delta \psi^I_\mu$. With the
solutions for $A$ and $g$, the gravitino variation gives
\begin{equation}
\frac{df}{dr}=-\frac{g_1^2 g_2^2  (e^{2 g_1 r}+1) (g_1^2 (e^{2 g_1
r}-1)+g_2^2 (e^{2 g_1 r}+1)^3)}{h_1 (g_2 e^{2 g_1 r}-g_1+g_2)^2 (g_2
e^{2 g_1 r}+g_1+g_2)^2}\, .
\end{equation}
After going to $\tilde{r}$ coordinate, we find the solution
\begin{equation}
f=g_1\tilde{r}-\ln [2(1+e^{2g_1\tilde{r}})]+\frac{1}{2}\ln
[2(g_1^2-g_2^2(1+e^{2g_1\tilde{r}})^2)]
\end{equation}
where, as usual, we have ignored all additive constants because they
can be absorbed by rescaling $x^0$ and $x^1$. The AdS$_3$ radius is
given by
\begin{equation}
L=\frac{e^{g_0}}{g_1}=\frac{1}{\sqrt{V_0}}.
\end{equation}
We can compute the ratio of the central charges between the two
fixed points
\begin{equation}
\frac{c_{\textrm{II}}}{c_{\textrm{III}}}=\frac{1}{1-\frac{g_1^2}{g_2^2}}>1.
\end{equation}
By the c-theorem, we see that point II and III correspond to the UV
and IR CFTs, respectively. This is in agreement with the solution
found in section \ref{6Dflow}. So, the solutions from both theories
are the same. This is the result, at the level of solutions, of the
fact that the two theories are equivalent as shown in \cite{KK}.
Near the UV point, the scalars behave as
\begin{equation}
\delta A \sim  e^{-2r/L},\qquad \delta h\sim e^{-4r/L},\,\,\,
L=\frac{h_1}{g_1^2}\, .
\end{equation}
We see that the flow is driven by a marginal operator dual to $A$ of
dimension 2. $h$ is dual to an irrelevant operator of dimension 4.
Up to quadratic order in the scalars, the potential
\eqref{potential40} at the UV point is given by
\begin{equation}
V=-\frac{g_1^4}{h_1^2}+\frac{4g_1^4}{h_1^2}h^2\, .
\end{equation}
We find that the scalar $A$ is massless at this point and dual to a
marginal operator. The scalar kinetic terms are
\begin{equation}
\mathcal{L}_{\textrm{scalar
kinetic}}=\frac{1}{2}\bigg(\frac{3A'^2}{(A^2-1)^2}+h'^2\bigg)\label{kinetic}.
\end{equation}
At the UV point, $A=0$, all the kinetic terms are canonically
normalized, and we can read off the values of mass squared directly
from the potential. In the unit of $\frac{1}{L^2}$, $h$ has
$m^2L^2=8$ which gives exactly $\Delta=4$ in agreement with the
asymptotic behavior. At the IR point, $A$ becomes massive with
positive mass squared as can be seen from the expansion of the
potential
\begin{equation}
V=-\frac{g_1^4}{h_1^2\big(1-\frac{g_1^2}{g_2^2}\big)}+\frac{12g_1^4g_2^8}{(g_1^2-g_2^2)^4h_1^2}A^2+\frac{4g_1^4g_2^4}{(g_1^2-g_2^2)^2h_1^2}h^2\,
.
\end{equation}
The positive mass squared means that the potential has a minimum at
the IR point, and the dual operator is irrelevant as it should. To
compute the mass squared at the IR point, we redefine $A$ to
$A=\tanh{\phi}$. Then, \eqref{kinetic} becomes
\begin{equation}
\frac{1}{2}(3\phi'^2+h'^2).
\end{equation}
The potential near the IR point is, to quadratic terms in scalars,
given by
\begin{equation}
\frac{V}{V_{0\textrm{IR}}}=1-4h^2-\frac{12g_2^4}{(g_1^2-g_2^2)^2}\phi^2.
\end{equation}
At this fixed point, $h$ and $\phi$ have $m^2L^2=8$ and
$m^2L^2=\frac{24}{1-\frac{g_1^2}{g_2^2}}$, respectively. We find
that
\begin{equation}
\Delta_{h}=4\qquad \textrm{and}\qquad
\Delta_{\phi}=1+\sqrt{1+\frac{24g_2^4}{(g_1^2-g_2^2)^2}}>2\, .
\end{equation}
As expected, the operator dual to $\phi$ is irrelevant with
dimension greater than 2.
%%%%%%%%%%%%%%%%%%%%%%%%%%%%%%%%%%%%%%%%%%%%%%%%%%%%%%%%%%%%%%%%%%%%%%%%%%%%%%%%%%%%%%%%%%%%%%%%%%%%%%%%%%%%%%%%%%%%%%%%%%%%%%%%%%%%%%%%%%%%%%%%
\section{Derivation of the central charges}\label{derivation}
In this Appendix we present a holographic derivation of the left- and right-
central charges in a 2D CFT which, to our knowledge, has not
appeared in the literature.
We follow \cite{Witten} and find the correlation functions of the
stress-energy tensor. The two point function will directly give the
value of the central charge. We first start with the gravitational
action in three dimensions of the form
\begin{equation}
I=\alpha \left[\int_M d^3x (\sqrt{G}R_G+2\Lambda)-\int _{\pd
M}d^2x\sqrt{g}2K\right] \, .\label{gravity action}
\end{equation}
The second term is the Gibbons-Hawking term with the induce metric
$g$ and extrinsic curvature of the boundary $\pd M$, $K$. The
coefficient $\alpha$ is dimensionless provided that we use the unit
$AdS_3$ in the measure $d^3x\sqrt{g}$. We then adopt Euclidean
signature and take the metric \cite{skenderis}
\begin{equation}
ds^2=G_{\mu\nu}dx^\mu dx^\nu
=\ell^2\left(\frac{d\rho^2}{4\rho^2}+\frac{g_{ij}(x,\rho)}{\rho}dx^idx^j\right).\label{3dmetric}
\end{equation}
In these coordinates, the extrinsic curvature is given by
$K_{ij}=-\frac{2\rho}{\ell}\pd_\rho g_{ij}$, and the boundary is at
$\rho=0$. We look for the solution of $g_{ij}$ of the form
\begin{equation}
g(x,\rho)=g_{(0)}+\rho g_{(2)}+h_{(2)}\rho \ln \rho+\ldots\,
.\label{g expansion}
\end{equation}
The Einstein equations give \cite{skenderis}
\begin{eqnarray}
\rho[2g''-2g'g^{-1}g'+\textrm{Tr}(g^{-1}g')g']+R_g-\textrm{Tr}(g^{-1}g')g&=&0\\
\nabla_i\textrm{Tr}(g^{-1}g')-\nabla^jg'_{ij}&=&0\\
\textrm{Tr}(g^{-1}g'')-\frac{1}{2}\textrm{Tr}(g^{-1}g'g^{-1}g')&=&0\,
.
\end{eqnarray}
Using the expansion \eqref{g expansion}, we find the relevant
equations
\begin{eqnarray}
\nabla^jg^{(2)}_{ij}&=&\frac{1}{2}\nabla_i R_g\label{g2eq}\\
\textrm{Tr}g_{(2)}&=&=\frac{1}{2}R_g\label{trace eq}\, .
\end{eqnarray}
We now expand the background metric $g_{(0)}$ about the flat metric.
We use the complex coordinates with the convention of
\cite{Polchinski} and write the metric $g_{(0)}$ as
\begin{equation}
g_{(0)}=\left(
          \begin{array}{cc}
            h_{zz} & \frac{1}{2}+h_{z\bar{z}} \\
            \frac{1}{2}+h_{z\bar{z}} & h_{\bar{z}\bar{z}} \\
          \end{array}
        \right).
\end{equation}
To simplify the equations, we will keep only the
$h_{\bar{z}\bar{z}}$ component non zero. This is enough to find the
$T_{zz}T_{zz}$ correlation functions. In the complex coordinates,
equation \eqref{g2eq} takes the form
\begin{equation}
\nabla_zg^{(2)}_{\bar{z}z}+\nabla_{\bar{z}}g^{(2)}_{zz}-2h_{\bar{z}\bar{z}}\nabla_zg^{(2)}_{zz}=\frac{1}{4}\pd_zR_g\,
. \label{g2zz}
\end{equation}
Equation \eqref{trace eq} gives
\begin{equation}
g_{z\bar{z}}^{(2)}-h_{\bar{z}\bar{z}}g^{(2)}_{zz}=\frac{1}{8}R_g\, .
\end{equation}
This can be used to eliminate $g^{(2)}_{z\bar{z}}$ in \eqref{g2zz}.
We finally find
\begin{equation}
\pd_{\bar{z}}g^{(2)}_{zz}-h_{\bar{z}\bar{z}}\pd_zg^{(2)}_{zz}-2\pd_zh_{\bar{z}\bar{z}}g^{(2)}_{zz}=\frac{1}{8}\pd_zR_g\,
.\label{master eq}
\end{equation}
This equation has precisely the structure of Virasoro's Ward
identity for the generating function of connected correlators of
$T_{zz}$. A different holographic derivation of it has been
discussed in \cite{banados}.
\\ \indent
The Ricci scalar $R_g$ is given by
\begin{equation}
R_g=2g_{(0)}^{z\bar{z}}R_{z\bar{z}}=4\pd_z^2h_{\bar{z}\bar{z}}\, .
\end{equation}
We can now solve \eqref{master eq} for $g^{(2)}$ order by order. The
first order equation is simply given by
\begin{equation}
\pd_{\bar{z}}g^{(2)}_{zz}=\frac{1}{2}\pd_z^3h_{\bar{z}\bar{z}}\, .
\end{equation}
This is easily solved by recalling $\pd_{\bar{z}}\frac{1}{z}=2\pi
\delta(z)$ and taking
\begin{equation}
g^{(2)}_{zz}=-\frac{3}{2\pi}\int
d^2w\frac{1}{(z-w)^4}h_{\bar{w}\bar{w}}(w)\, .\label{first order
sol}
\end{equation}
Back to our action \eqref{gravity action}, we can evaluate this
action on the solution \eqref{3dmetric} with the expansion \eqref{g
expansion}. This gives \cite{kraus}
\begin{equation}
\delta I=\alpha \int d^2x\sqrt{g}
(g^{(2)ij}-g_{(0)}^{kl}g^{(2)}_{kl}g^{(0)ij})\delta g^{(0)}_{i,j}\,
.
\end{equation}
Although, in our coordinates, the boundary is at the lower limit of
the $\rho$ integration, $\rho=0$, in contrast to \cite{kraus} in
which the boundary is at the upper limit of the $\eta$ integration,
$\eta=\infty$, we find $\delta I$ with the same sign as that in
\cite{kraus}. This is because of the extra minus sign in the
extrinsic curvature $K_{ij}$.
\\
\indent In the complex coordinates and with only
$h_{\bar{z}\bar{z}}\neq 0$, we find
\begin{equation}
\delta I=2\alpha i\int d^2z g^{(2)}_{zz}\delta
g^{(0)}_{\bar{z}\bar{z}}\, .
\end{equation}
This gives the one point function for the stress-energy tensor.
Using the solution \eqref{first order sol}, we can find the two
point function by differentiate one more. The result is
\begin{equation}
\langle T(z)T(w)\rangle=(-2\pi)^2\frac{\delta^2}{\delta
h_{\bar{z}\bar{z}}(z)\delta h_{\bar{w}\bar{w}}(w)}
e^{iS}|_{h_{\bar{z}\bar{z}}=0}=\frac{12\pi\alpha}{(z-w)^4}\,
.\label{TTcorrelator}
\end{equation}
We have used our normalization factor of $-2\pi$ in the definition
of the stress-energy tensor. This normalization has been determined
by computing the three point function $\langle
T(z_1)T(z_2)T(z_3)\rangle$ which in turn can be obtained by solving
\eqref{master eq} to the second order. After matching this three
point function with the CFT $\langle T(z_1)T(z_2)T(z_3)\rangle$, we
find the normalization factor. We then compare \eqref{TTcorrelator}
with the OPE $T(z)T(w)\sim \frac{c_L}{2(z-w)^4}+\ldots$, we obtain
\begin{equation}
c_L=24\pi \alpha\, .
\end{equation}
A similar analysis can be done for the $\langle
\bar{T}\bar{T}\rangle$ with non zero $h_{zz}$. The right moving
central charge is then given by
\begin{equation}
c_R=24\pi \alpha\, .
\end{equation}
\indent In principle, we can use \eqref{master eq} to find any $n$
point function of the CFT's stress-energy tensor. However, the above
analysis only involves ether $h_{\bar{z}\bar{z}}$ or $h_{zz}$. With
all $h_{zz}$, $h_{z\bar{z}}$ and $h_{\bar{z}\bar{z}}$ non zero, we
have also checked, to leading order, that there is no $T\bar{T}$
correlation function, but there is a coupling between $h_{z\bar{z}}$
and $h_{zz}$ and between $h_{z\bar{z}}$ and $h_{\bar{z}\bar{z}}$.
These couplings can be removed by adding some local counter terms to
the two dimensional action. Beyond leading order, it is not clear
what we can learn from
a very complicated equation coming from \eqref{master eq}. \\
\indent We end this section by briefly discussing the contribution
to the Kac-Moody level from the gauge Chern-Simons term. Following
\cite{kraus partition}, the gauge field can be expanded as
\begin{equation}
A=A^{(0)}+\rho A^{(1)}+\ldots\, .
\end{equation}
The Lagrangian for the gauge field including the Chern-Simons term
is
\begin{equation}
I=-\frac{1}{2}\int*F^a\wedge F^a-\frac{\beta}{2} \int
\left(A^a\wedge dA^a+\frac{2}{3}f_{abc}A^a\wedge A^b\wedge
A^c\right).
\end{equation}
We will suppress the gauge group index on $A$ from now on to make
the expression compact. From this action, it is straightforward to
find the equation of motion and find, in the $A_\rho=0$ gauge,
\begin{equation}
\pd_iA^{(0)}_j-\pd_jA^{(0)}_i=0\, .\label{flatness}
\end{equation}
As discussed in \cite{kraus partition}, it is necessary to add a
boundary term in order to obtain only the left moving current. This
boundary term is given by
\begin{equation}
I_b=\frac{\beta}{2}\int d^2x \sqrt{g}g^{ij}A^{(0)}_i A^{(0)}_j\, .
\end{equation}
Notice the sign change as oppose to the result in \cite{kraus
partition}. We can solve \eqref{flatness} in complex coordinates by
taking
\begin{equation}
A_z(z)=-\frac{1}{2\pi}\int d^2w\frac{1}{(z-w)^2}A_{\bar{w}}(w)\, .
\end{equation}
Inserting this into $I_b$, we obtain
\begin{equation}
I_b=-i\frac{\beta}{2\pi}\int d^2z
d^2wA_{\bar{w}}(w)\frac{1}{(z-w)^2}A_{\bar{z}}(z)\, .
\end{equation}
We can now find
\begin{equation}
\langle J(z)J(w)\rangle=(-2\pi)^2\frac{\delta^2}{\delta
A_{\bar{z}}(z)\delta
A_{\bar{w}}(w)}e^{iI_b}|_{A_{\bar{z}}=0}=\frac{4\pi \beta}{(z-w)^2}
\end{equation}
which gives $k=8\pi\beta$ by using the OPE $J(z)J(w)\sim
\frac{k}{2(z-w)^2}+\ldots$. The factor $-2\pi$ is again due to our
normalization of the current. The central charge is
$c=6k=48\pi\beta$.
%%%%%%%%%%%%%%%%%%%%%%%%%%%%%%%%%%%%%%%%%%%%%%%%%%%%%%%%%%%%%%%%%%%%%%%%%%%%%%%%%%%%%%%%%%%%%%%%%%%%%%%%%%%%%%%%%%%%%%%%%%%%%%%%

\end{document}